\documentclass{emulateapj}
\shorttitle{ROTSE-III Observations of GRB 030329}
\shortauthors{Smith et al.}
\newcommand{\PSbox}[3]{\mbox{\includegraphics{#1}\hspace{#2}\rule{0pt}{#3}}}

\begin{document}

\slugcomment{Accepted to Astrophysical Journal Letters, 4 Sept 2003}
 
\title{ROTSE-III Observations of the Early Afterglow From GRB 030329}

\author{Smith, D. A.\altaffilmark{1}, Rykoff, E. S.\altaffilmark{1}, Akerlof,
	C. W.\altaffilmark{1}, Ashley, M. C. B.\altaffilmark{2}, Bizyaev,
	D.\altaffilmark{4,6}, McKay, T. A.\altaffilmark{1}, Mukadum,
	A.\altaffilmark{5}, Phillips, A.\altaffilmark{2}, Quimby,
	R.\altaffilmark{5}, Schaefer, B.\altaffilmark{5}, Sullivan,
	D.\altaffilmark{7}, Swan, H. F.\altaffilmark{1}, Vestrand,
	W. T.\altaffilmark{3}, Wheeler, J. C.\altaffilmark{5}, and Wren,
	J.\altaffilmark{3}}

\email{donaldas@umich.edu}

\altaffiltext{1}{University of Michigan, 2477 Randall Laboratory,
	500 E. University Ave., Ann Arbor, MI, 48104}
\altaffiltext{2}{School of Physics, Department of Astrophysics and Optics, University of New South Wales, Sydney, NSW 2052, Australia}
\altaffiltext{3}{Los Alamos National Laboratory, NIS-2 MS D436, Los Alamos, NM 87545}
\altaffiltext{4}{Department of Physics, University of Texas at El Paso, El Paso, Texas, USA}
\altaffiltext{5}{Department of Astronomy, University of Texas, Austin, TX 78712}
\altaffiltext{6}{Sternberg Astronomical Institute, Moscow, Russia}
\altaffiltext{7}{School of Chemical \& Physical Sciences, Victoria University of Wellington, PO Box 600, Wellington, NZ}

\begin{abstract}

Using two identical telescopes at widely separated longitudes, the ROTSE-III
network observed decaying emission from the remarkably bright afterglow of
GRB~030329. In this report we present observations covering 56\% of the period
from 1.5--47~hours after the burst.  We find that the light curve is piecewise
consistent with a powerlaw decay.  When the ROTSE-III data are combined with
data reported by other groups, there is evidence for five breaks within the
first 20~hours after the burst.  Between two of those breaks, observations from
15.9--17.1~h after the burst at 1-s time resolution with McDonald Observatory's
2.1-m telescope reveal no evidence for fluctuations or deviations from a simple
power law.  Multiple breaks may indicate complex structure in the jet.  There
are also two unambiguous episodes at 23 and 45~hours after the burst where the
intensity becomes consistent with a constant for several hours, perhaps
indicating multiple injections of energy into the GRB/afterglow system.

\end{abstract}
\keywords{Gamma-rays: bursts}

\section{Introduction}
 
The ROTSE program is dedicated to recording optical observations of Gamma-Ray
Bursts (GRBs) within seconds of their detection at high energies.  Although the
study of late-time afterglows has revolutionized our understanding of GRBs,
comparatively little is known about the nature and diversity in the prompt
optical and early afterglow emission~\citep{piran99}.  To date the ROTSE-I
observations of GRB~990123 remain the only optical detections of a GRB while
the burst was still bright in gamma-rays~\citep{abbbb99}.  ROTSE-I and LOTIS
observations of many bursts demonstrate that prompt emission is not, in
general, well represented by extrapolation of late time decay
curves~\citep{kabbb01,ppwab99}.  More sensitive instruments are required to
study prompt emission from typical bursts.

In preparation for HETE-2~\citep{vvdjl99} and the Swift
satellites~\citep{gehre00} which can distribute rapid localizations of GRBs
with arcminute accuracy, the ROTSE team developed a new set of 0.45-m
telescopes.  The ROTSE-III instruments use fast optics (f/1.9) to yield a
$1\arcdeg.85 \times 1\arcdeg.85$ field of view over a four megapixel CCD
camera.  Four of these robotically controlled, automatically scheduled
instruments are being erected around the globe, both to increase the amount of
sky available for instant viewing and to enable around-the-clock coverage of
GRB afterglow behavior~\citep{akmrs03}.  At the time of this writing, the first
two instruments, labeled ROTSE-IIIa and ROTSE-IIIb, are operating at Siding
Spring Observatory in New South Wales, Australia, and McDonald Observatory at
Ft. Davis, Texas, USA, respectively.  The installation of ROTSE-IIIc is nearing
completion in Namibia, and ROTSE-IIId is slated to arrive in Turkey in the Fall
of 2003.

Within a week of the first two ROTSE-III instruments becoming fully
operational, the network was tested by the discovery with HETE-2 of GRB~030329.
Spectroscopic observations found the source to lie at a redshift of
$z=0.168$~\citep{gpekv03}, which makes it the closest GRB to the Earth yet to
be measured, with the possible exception of the anomalously faint
GRB~980425~\citep{gvpka99}.  This event provided the first direct spectral
evidence for a supernovae associated with a high redshift
GRB~\citep{smgmb03,hsmfw03,kdwmn03}.  We report here on the light curve of
GRB~030329 for the first two days of observations by ROTSE-IIIa \& b.

\begin{figure*}
\PSbox{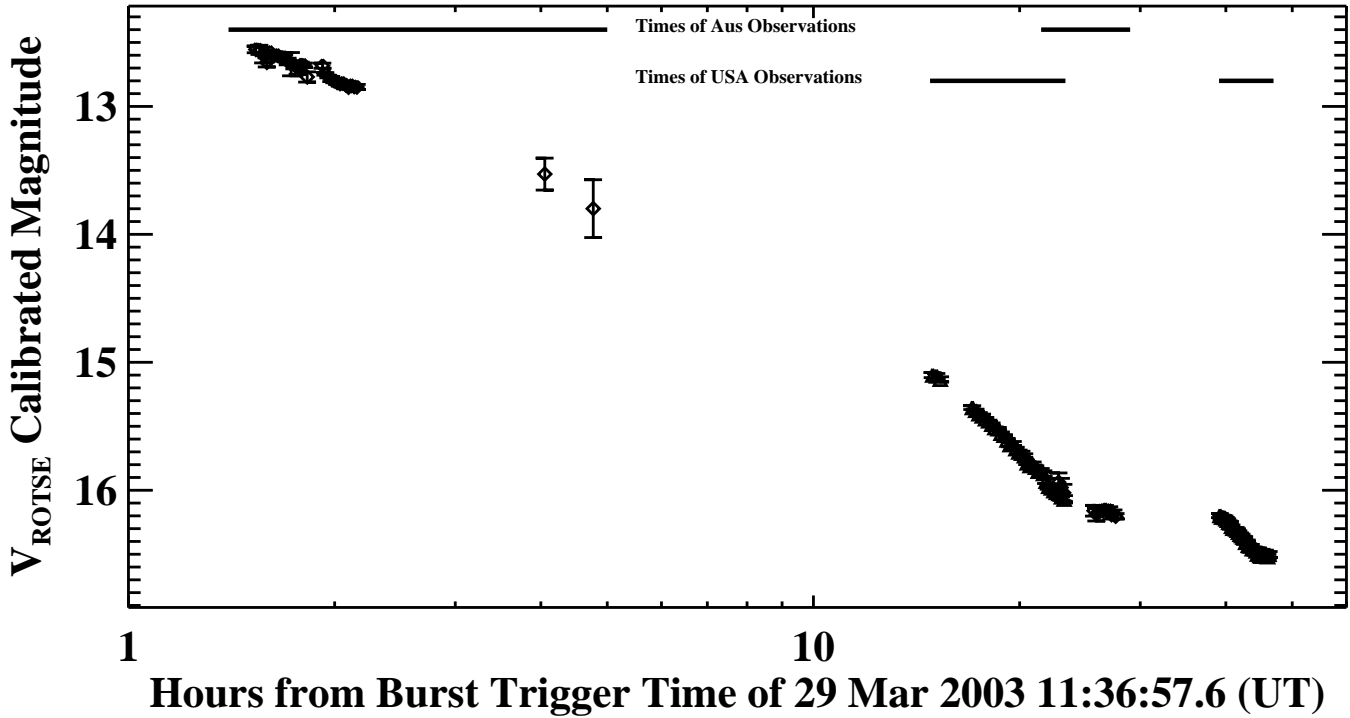 hoffset=-10 hscale=75 vscale=75 voffset=10}{6.0in}{3.9in}
\figcaption{Decay of GRB 030329 over the first two days of observations.
Magnitudes are calibrated to $V_{\rm ROTSE}$ as described in the text.
Diamonds indicate measurements made with ROTSE-IIIa and triangles those made
with ROTSE-IIIb.  Intervals when the source was above the horizon at each site
are indicated by horizontal bars at the top of the graph.\label{fig:ful}}
\end{figure*}

\section{Observations and Analysis}

GRB 030329 was first observed by the instruments on board the HETE-2 satellite
at 11:37:14.67 UTC.  Ground analysis of the SXC data produced a localization
that was reported in a Gamma-Ray Burst Coordinate Network (GCN) Notice 73
minutes after the burst~\citep{vcdvm03}.  Although the burst's celestial
location was above the Australian horizon at the time, the ROTSE-IIIa
connection to the GCN was temporarily experiencing sporadic outages.  It was
therefore impossible for the ROTSE automated scheduler daemon to respond to the
alert.  However, electronic mail was also distributed to ROTSE team members,
and a manual response by the ROTSE-IIIa telescope was initiated over the
Internet within 15~minutes of the alert's distribution, 88~minutes after the
burst itself.

Clouds and intermittent rain interfered with observations and delayed analysis
of the images, but once~\citet{petpr03} and~\citet{torii03} identified the
optical counterpart, we were able to extract a light curve from the setting
source.  Our earliest useful images began 92~minutes after the burst
event~\citep{ryksm03}.  At 4.8~h after the burst, the air mass to the source
grew too large for further observation.

\begin{figure}
\PSbox{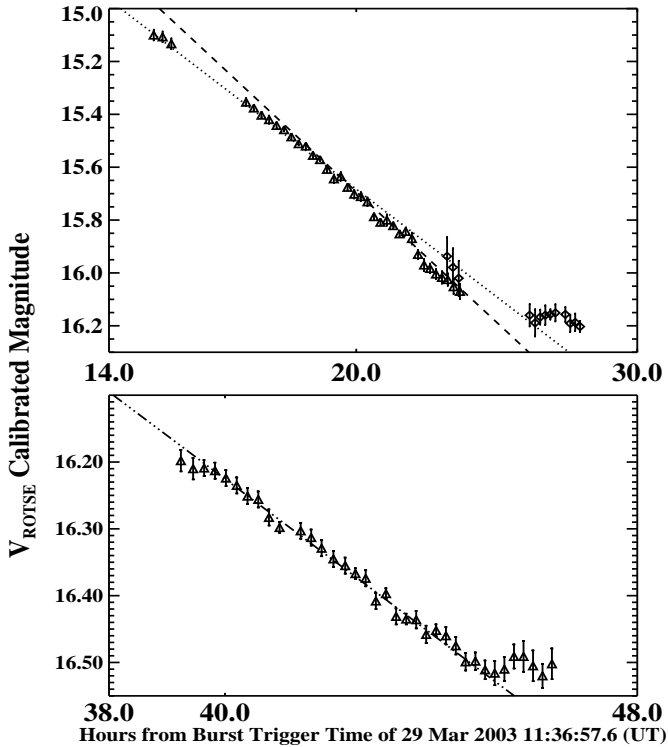 hoffset=0 hscale=50 vscale=45 voffset=-25}{3.0in}{4in}
\figcaption{Expanded views of two segments of the ROTSE-III light curve for
GRB~030329.  Diamonds indicate measurements made with ROTSE-IIIa and triangles
those made with ROTSE-IIIb.  The top panel shows the interval from 14 to 30~h
after the burst, while the bottom panel covers the time from 38 to 48~h.  These
panels also overlay best-fit power law functions onto the observations.  The
data in the top panel are clearly inconsistent with a single power law decay.
Parameters for these functions are given in Table~\protect{\ref{tab:pars}} The
burst intensity levels off at $\sim23$~h after the burst, maintaining a
constant intensity for the rest of the Australian night, but begins decaying
again in the second night of Texas observations, as shown in the bottom panel,
before entering a second flattening episode.
\label{fig:sub}}
\end{figure}

ROTSE-IIIb (in Texas) began observations 14.8~h after the burst, and followed
the source for a further 8.5~h.  Once the source had risen again in Australia,
contemporaneous observation from both ROTSE-III instruments was possible for
1.8~h.  The ROTSE-III automatic scheduler normally plans observations of fading
GRB afterglows to occur at logarithmically increasing intervals, leading to
gaps in coverage at the start of each night's observing.  As it became clear
how bright this source was, we overrode the default schedule to instruct the
instruments to look at nothing else as long as GRB~030329 was above the local
horizon.

The ROTSE-III images were bias-subtracted and flat-fielded in the standard way.
For images recorded later than 10~h after the burst, we co-added sets of ten
frames before deriving a light curve.  We applied SExtractor~\citep{berar96}
with aperture photometry to these co-added images to detect objects and derive
their intensities.  We compared these intensities to the USNO~A2.0 catalog to
derive an approximate R-band magnitude zero-point offset for the field.
Zero-point offsets among the ROTSE images are determined through application of
a relative photometry procedure. Absolute calibration to a V-magnitude
equivalent (dubbed $V_{\rm ROTSE}$) is derived by comparison of several nearby
stable stars to the $UBVR_cI_c$ photometry of~\citet{h2023}.

The offset between unfiltered $V_{\rm ROTSE}$ observations and standard V
magnitudes is a function of color. Without color information for the early-time
afterglow, we have not applied any color correction to these data. However,
from 6.0--14.4~h after the burst, during observations reported
by~\citet{bspdt03}, the afterglow color is rather blue, with $V-R\approx0.3$.
This color remained stable through a break in the lightcurve.  If the afterglow
maintained this color throughout the ROTSE-III observations, it would be
appropriate to further adjust the $V_{\rm ROTSE}$ magnitudes reported here to
be approximately 0.5 magnitudes brighter.

GRB~030329 was also observed with a time resolution of 1-second using a CCD
time-series photometer mounted at the prime focus position of the McDonald
Observatory 2.1-m telescope.  The photometer ``Argos''~(Nather \&~Mukadam, in
preparation) operates in frame transfer mode, and thus allows rapid exposures
as short as one second.  The time-series photometry started 15.87 hours after
the burst trigger, and continued for 4290 seconds in essentially photometric
conditions.  No filter was used, so the spectral response of the instrument was
dominated by the combination of a typical CCD sensitivity and the Earth's
atmosphere: the transmission is about 70\% in the wavelength range 560--640~nm,
and greater than 50\% in the wavelength range 450--800~nm.

Typical ADU count rates for the optical transient were 17,000 per second, while
the sky contributed about 10,500 counts per second in the chosen software
aperture.  There were two comparison stars in the 2.8~arcminute field of view
of the instrument, and sky-subtracted light curves for each object were
obtained by using an optimized circular software aperture.  A corrected light
curve for the target object was then generated by dividing its light curve to
the sum of the two comparison star light curves (which have been normalized to
unity).  

\section{Results}\label{sec:results}

The full light curve from both ROTSE-IIIa \& b is shown on a logarithmic scale
in the top panel of Figure~\ref{fig:ful}.  Figure~\ref{fig:sub} shows two
segments of the light curve on expanded scales.  In this section, we describe
the behavior of the afterglow during each observing sequence.

The poor observing conditions during the first observing run (from 1.5 to 4.8~h
after the burst) introduce large systematic errors into the photometry, but
there is no evidence for any significant deviations from a power law decay with
an index of 0.88$\pm$0.05, a value not atypical for early GRB afterglows.  The
earliest detections reveal an unfiltered $V_{\rm ROTSE}$~magnitude of 12.55,
over 100 times brighter than any other afterglow to date to be observed at this
epoch~\citep{pfkps03}, consistent with the very low redshift for the object.
The latest points in this sequence, recorded after an hour during which
ROTSE-IIIa automatically ceased operation due to a light rain, are also
consistent with the same decay curve.

When ROTSE-IIIb commenced observations at 14.8~h, the intensity of GRB~030329
was decaying much more rapidly, consistent with a power law index of
1.85$\pm$0.07 and strongly inconsistent with the earlier curve observed with
ROTSE-IIIa.  If we assume a single, sharp transition from one decay mode to the
other, these data predict a transition time of $\sim$12--14~h after the burst.
Over the course of the 8.5~h observing sequence with ROTSE-IIIb on 30 Mar 2003,
the decay slope steepened further, demanding the introduction of a third power
law with decay index of 2.25$\pm$0.08.  The best-fit transition time to this
decay slope is at 19.0~h after the burst.  

The interval during which GRB~030329 was observed by the 2.1-m coincided with
the early part of this decay phase, before the break at 19~h.  The corrected
light curve at 1-s resolution shows the optical transient to be fading
smoothly.  The fitted power law decline has an index of 1.98$\pm$0.02. This
power law fits well to both the beginning and ending 500-second time intervals
with the same normalization to within 0.2\%, thus indicating that there were no
departures from the overall power law decline.  The largest deviation from the
best fit power law was one point that was low by only $4.4\sigma$.  Analysis of
binned light curves reveals no significant fluctuations from the power law on
any longer time scales.  The computed discrete Fourier transform did not reveal
any power out to the Nyquist frequency of 0.5~Hz up to a limit of 2~millimag
(i.e., 0.2\% change in intensity).  In all, we found no deviations from a
simple power law decline.

During the interval when both ROTSE-III instruments could observe the source
simultaneously, our calibration procedure yielded consistent results.  However,
ROTSE-IIIa subsequently observed the source intensity to level off and remain
constant for at least three hours.  When the source again became visible to
ROTSE-IIIb in Texas, it was found to still have the same intensity as when last
seen by ROTSE-IIIa.  It shortly commenced a rapid decay phase, consistent with
a single power law of index 2.34$\pm$0.08.  This decay rate is consistent with
the decay slope observed at the end of the previous night.  The decay phase
lasted for approximately six hours.  The source then entered a second
flattening phase that was still continuing when observations ceased for the
night two hours later\footnote{The ROTSE-III light curve is stored in
electronic form at http://www.rotse.net/transients/grb030329}.

The decay curve for GRB~030329 over its first two days is clearly inconsistent
with a single power-law function.  We find that at least four power laws plus
two intervals of constant intensity are necessary to adequately reproduce the
behavior of the light curve.  Three of these functions are plotted as broken
lines in Figure~\ref{fig:sub}.  The decay indices as well as the intervals over
which each power-law segment applies are given in Table~\ref{tab:pars}.

\section{Discussion}

The world-wide distribution of the ROTSE-III network allows for unprecedented
coverage of the early decay curve of GRB afterglows, and in the case of
GRB~030329, the afterglow was above the horizon for 56\% of the first 47~h
after the event.  Since the ROTSE-III network was only half complete at the
time of GRB~030329, there are still large gaps in the derived light curve when
the source was not visible to either telescope.  However, at least two groups
have published observations of the afterglow's behavior during the time on the
first day when neither ROTSE-IIIa nor ROTSE-IIIb could observe it.

\citet{ukiym03} observed GRB~030329 with a variety of telescopes from 1.2~h to
$\sim14$~h after the event, and they report three breaks in the light curve
during this time: from 0.74$\pm$0.02 to 0.95$\pm$0.01 at $t=2.04$~h, to
0.65$\pm$0.04 at $t=3.9$~h, and again to 1.16$\pm$0.01 at $t=5.4$~h.  This
characterization is consistent with the ROTSE-IIIa data, which have neither
sufficient coverage nor small enough errors to distinguish between this and the
best-fit single power law reported above in Section~\ref{sec:results}.
\citet{bspdt03} present observations taken with the 1.5~m Russian-Turkish
Telescope shortly after the time of this last transition and give a slope of
1.19$\pm$0.01 from 6.2~h to 14.4~h after the burst, after which the slope
changes to $\sim1.9$, which is consistent with the ROTSE-IIIb observations at
14.8~h.  The Japanese and Turkish programs, therefore, neatly fill in the gap
between ROTSE-III observations, and the joint data set provides evidence for
five changes in slope decay index within the first 20~hours after the burst.

\begin{deluxetable}{lll}
\tablecaption{Parameters of Best-Fit Power-Law Decay Curve\label{tab:pars}}
\tablehead{
\colhead{Start Time (h)}  & \colhead{Stop Time (h)} & \colhead{Index}
}
\startdata
1.5 & 4.8 & 0.87$\pm$0.03 \\
14.8 & 19.0 & 1.85$\pm$0.06 \\
19.0 & 23.0 & 2.25$\pm$0.08 \\
23.0 & 26.0 & $<$0.4 \\
39.5 & 45.0 & 2.34$\pm$0.08 \\ 
45.0 & 46.5 & $<$0.4 
\enddata
\end{deluxetable}

No GRB afterglow has been observed nearly as intensely as GRB~030329, and for
most other bursts only a single steepening in the light curve decay has been
reported.  These breaks have been taken as evidence for deceleration of a
relativistic jet: as the jet decelerates, the beaming angle must increase, and
when the beaming angle exceeds the jet angle, the intensity of visible light is
expected to decrease more rapidly~\citep{rhoad97,sapih99,psf03}.  Multiple
breaks within a single decay curve are more difficult to interpret in this
paradigm, but \citet{weiji02}, for example, have suggested that they may result
from off-axis viewing geometry or perhaps non-uniform structure within the jet.
A detailed analysis of what geometry five decay breaks might demand is beyond
the scope of this paper, but the data clearly are not consistent with the
simplest models and demand caution when trying to interpret $t\sim0.5$~d as
{\it the} jet break~\citep{gnp03,bspdt03,pfkps03}.

The other striking feature of the afterglow is the deviation from its rapid
decay, beginning at $\sim23$~h after the burst.  In the ROTSE-III observations,
the intensity levels off and remains constant for several hours.  A second
decay phase begins $\sim39$~h after the burst.  It is unlikely that this
episode is related to the emergence of a supernova in the light curve, as the
first spectral evidence for a supernova associated with GRB~030329 emerged only
after $\sim7$~d~\citep{smgmb03}.  At one day after the event, the optical
intensity should be dominated by the burst afterglow.  Furthermore, the light
curve flattens yet again at $\sim44$~h.  \citet{weiji02} argue that geometric
factors could produce a flattening of the light curve at about this time, but
it is hard to understand the repeating nature of the phenomenon within that
paradigm.  On the other hand, \citet{gnp03} have interpreted these episodes as
refreshed shocks produced by further ejection events from the central engine
catching up with the decelerating blast wave.  In this interpretation, the
central engine remains an active part of the GRB process well into the
afterglow phase, further underscoring the need for early and thorough
observations of the entire burst event if the complex physical interactions are
ever to be teased out.

\acknowledgments 

This work has been supported by NASA grants NAG 5-5281 and F006794, NSF grant
AST 01-19685, the Michigan Space Grant Consortium, the Australian Research
Council, the University of New South Wales, the University of Texas, and the
University of Michigan. DAS is supported by an NSF Astronomy and Astrophysics
Postdoctoral Fellowship under award AST 01-05221.  Work performed at LANL is
supported by NASA SR\&T through Department of Energy (DOE) contract
W-7405-ENG-36 and through internal LDRD funding.

\newcommand{\noopsort}[1]{} \newcommand{\printfirst}[2]{#1}
  \newcommand{\singleletter}[1]{#1} \newcommand{\switchargs}[2]{#2#1}


\begin{thebibliography}{}

\bibitem[\protect\astroncite{{Akerlof} {\rm et~al.\/}}{1999}]{abbbb99}
 {Akerlof}, C., et~al.
\newblock 1999, \nat, 398, 400

\bibitem[\protect\astroncite{{Akerlof} {\rm et~al.\/}}{2003}]{akmrs03}
 {Akerlof}, C., et~al.
\newblock 2003, \pasp, 115, 132

\bibitem[\protect\astroncite{{Bertin} \& {Arnouts}}{1996}]{berar96}
{Bertin}, E. \& {Arnouts}, S.
\newblock 1996, \aaps, 117, 393

\bibitem[\protect\astroncite{{Burenin} {\rm et~al.\/}}{2003}]{bspdt03}
 {Burenin}, R., et~al.
\newblock 2003, Astronomy Letters, 9, 1, astro-ph/0306137

\bibitem[\protect\astroncite{{Galama} {\rm et~al.\/}}{1999}]{gvpka99}
 {Galama}, T.~J., et~al.
\newblock 1999, \aaps, 138, 465

\bibitem[\protect\astroncite{{Gehrels}}{2000}]{gehre00}
{Gehrels}, N.~A. 2000, , in Proc. SPIE Vol. 4140, p. 42-49, X-Ray and Gamma-Ray
  Instrumentation for Astronomy XI, Kathryn A. Flanagan; Oswald H. Siegmund;
  Eds., 42

\bibitem[\protect\astroncite{{Granot}, {Nakar}, \& {Piran}}{2003}]{gnp03}
{Granot}, J., {Nakar}, E., \& {Piran}, T.
\newblock 2003, astro-ph/0304563

\bibitem[\protect\astroncite{Greiner {\rm et~al.\/}}{2003}]{gpekv03}
Greiner, J., Peimbert, M., Estaban, C., Kaufer, A., Vreeswijk, P., Jaunsen, A.,
  Smoke, J., Klose, S., \& Reimer, O.
\newblock 2003, GCN Circ. No. 2020

\bibitem[\protect\astroncite{Henden}{2003}]{h2023}
Henden, A.
\newblock 2003, GCN Circ. No. 2023

\bibitem[\protect\astroncite{{Hjorth} {\rm et~al.\/}}{2003}]{hsmfw03}
{Hjorth}, J. et~al.
\newblock 2003, \nat, 423, 847

\bibitem[\protect\astroncite{{Kawabata} {\rm et~al.\/}}{2003}]{kdwmn03}
{Kawabata}, K.~S. et~al.
\newblock 2003, \apjl, in press, astro-ph/0306155

\bibitem[\protect\astroncite{{Kehoe} {\rm et~al.\/}}{2001}]{kabbb01}
 {Kehoe}, R., et~al.
\newblock 2001, \apjl, 554, L159

\bibitem[\protect\astroncite{{Park} {\rm et~al.\/}}{1999}]{ppwab99}
 {Park}, H.~S., et~al.
\newblock 1999, \aaps, 138, 577

\bibitem[\protect\astroncite{{Perna}, {Sari}, \& {Frail}}{2003}]{psf03}
{Perna}, R., {Sari}, R., \& {Frail}, D.
\newblock 2003, \apj, astro-ph/0305145

\bibitem[\protect\astroncite{Peterson \& Price}{2003}]{petpr03}
Peterson, B. \& Price, P.
\newblock 2003, GCN Circ. No. 1985

\bibitem[\protect\astroncite{Piran}{1999}]{piran99}
Piran, T.
\newblock 1999, \physrep, 314(6), 575

\bibitem[\protect\astroncite{{Price} {\rm et~al.\/}}{2003}]{pfkps03}
 {Price}, P., et~al.
\newblock 2003, \nat, 423, 844

\bibitem[\protect\astroncite{{Rhoads}}{1997}]{rhoad97}
{Rhoads}, J.~E.
\newblock 1997, \apjl, 487, L1

\bibitem[\protect\astroncite{Rykoff \& Smith}{2003}]{ryksm03}
Rykoff, E.~S. \& Smith, D.~A.
\newblock 2003, GCN Circ. No. 1995

\bibitem[\protect\astroncite{{Sari}, {Piran}, \& {Halpern}}{1999}]{sapih99}
{Sari}, R., {Piran}, T., \& {Halpern}, J.~P.
\newblock 1999, \apjl, 519, L17

\bibitem[\protect\astroncite{{Stanek} {\rm et~al.\/}}{2003}]{smgmb03}
 {Stanek}, K.~Z., et~al.
\newblock 2003, \apjl, 591, L17

\bibitem[\protect\astroncite{Torii}{2003}]{torii03}
Torii, K.
\newblock 2003, GCN Circ. No. 1986

\bibitem[\protect\astroncite{{Uemura} {\rm et~al.\/}}{2003}]{ukiym03}
{Uemura}, M., {Kato}, T., {Ishioka}, R., {Yamaoka}, H., {Monard}, B., {Nogami},
  D., {Maehara}, H., {Sugie}, A., \& {Takahashi}, S.
\newblock 2003, \nat, astro-ph/0306396

\bibitem[\protect\astroncite{{Vanderspek} {\rm et~al.\/}}{2003}]{vcdvm03}
 {Vanderspek}, R., et~al.
\newblock 2003, GCN Circ. No. 1997

\bibitem[\protect\astroncite{{Vanderspek} {\rm et~al.\/}}{1999}]{vvdjl99}
{Vanderspek}, R., {Villase{\~ n}or}, J., {Doty}, J., {Jernigan}, J.~G.,
  {Levine}, A., {Monnelly}, G., \& {Ricker}, G.~R.
\newblock 1999, \aaps, 138, 565

\bibitem[\protect\astroncite{{Wei} \& {Jin}}{2003}]{weiji02}
{Wei}, D.~M. \& {Jin}, Z.~P.
\newblock 2003, \aap, 400, 415

\end{thebibliography}
\end{document}